\documentclass[12pt]{article}
\usepackage{epsfig}
\usepackage{amsmath}

\makeatletter
\@addtoreset{equation}{section}

\newcommand{\p}{\partial}

\begin{document}


\begin{titlepage}
\begin{center}

\hfill\parbox{4cm}{
{\normalsize\tt hep-th/0211090}\\
{\normalsize UT-Komaba/02-14}
}

\vskip 1in

{\LARGE
\bf S-brane Actions}

\vskip 0.3in

{\large Koji {\sc Hashimoto}$^a$\footnote{\tt
koji@hep1.c.u-tokyo.ac.jp},
Pei-Ming {\sc Ho}$^b$\footnote{{\tt
 pmho@phys.ntu.edu.tw}}
and John E.\ {\sc Wang}$^{b,c}$\footnote{{\tt
hllywd2@phys.ntu.edu.tw}} }

\vskip 0.15in

${}^a$ {\it Institute of Physics, University of Tokyo, Komaba}\\
{\it Tokyo 153-8902, Japan}\\[3pt]
${}^b$ {\it Department of Physics, National Taiwan University,
Taipei, Taiwan}\\
${}^c$ {\it Institute of Physics, Academia Sinica, Taipei, Taiwan}
\\[0.3in]

{\normalsize November, 2002}

\end{center}

\vskip .3in

\begin{abstract}
\normalsize\noindent We derive effective actions for Spacelike
branes (S-branes) and find a solution describing the formation of
fundamental strings in the rolling tachyon background.  The
S-brane action is a Dirac-Born-Infeld action for Euclidean
worldvolumes defined in the context of time-dependent tachyon
condensation of non-BPS branes.  It includes gauge fields and in
particular a scalar field associated with translation along the
time direction. We show that the BIon spike solutions constructed
in this system correspond to the production of a confined
electric flux tube (a fundamental string) at late time of the
rolling tachyon.
\end{abstract}

\vfill

\end{titlepage}
\setcounter{footnote}{0}

\pagebreak
\renewcommand{\thepage}{\arabic{page}}
\pagebreak


\section{Introduction}

The exploration of time-dependent backgrounds is of theoretical
interest and has cosmological applications.  In the context of
string theory a natural candidate to study is the tachyon
condensation process on unstable branes.  In investigating this
direction, techniques are being developed and in particular a new
type of Spacelike brane (S-brane)\cite{stro} has appeared. This
brane is defined as a time-dependent classical solution of the
tachyon system. While S-branes naturally arise in the tachyon
system especially with rolling tachyons\cite{roll}, their role and
physical significance in string theory, especially in
time-dependent backgrounds and dS/CFT, is still being developed
(see Ref.\ \cite{sonogo} for example). In this letter we derive an
effective action for S-branes, and show that the S-brane approach
can be successfully applied to show that fundamental strings
appear during tachyon condensation.


\section{Derivation of the S-brane action}

The form of general tachyon actions has been shown \cite{tsey} to
guarantee that spatial tachyon kinks are universally governed by
the Dirac-Born-Infeld (DBI) form of D-brane actions. We similarly
derive the effective action for time-dependent
tachyon kinks representing S-branes
and show it has a universal form so essential features of the
S-brane action are independent of the explicit tachyon action.
Examining the tachyon system of a non-BPS D$(p\!+\!1)$-brane we
now construct time-dependent classical solutions representing
S$p$-branes. Using the action
\begin{eqnarray}
  S = -\int d^{p+2}x \; V(T) \sqrt{1 +
(\partial_\mu T)^2} \label{senac}
\end{eqnarray}
originally proposed in Ref.\ \cite{tact}, we will see how the rolling
tachyon picture in Ref.\ \cite{roll} is reconciled with the S-brane
picture.  Here $T(x^\mu)$ with $ \mu = 0,1,\cdots,p\!+\!1$ is the
tachyon field and dot will denote a derivative with respect to
time $x^0$. The tachyon potential achieves its maximum at $T=0$
and asymptotes to zero (closed string vacuum) at large $T$.
This effective action gives
the known exponentially decreasing pressure at late times while
being consistent with the string theory calculation where $V(T)$
is taken to be an exponential function of $T$.
For simplicity we take the tachyon classical solution to depend
only on time $x^0$.  Using the fact that energy $ {\cal E} =
V(T)/\sqrt{1-\dot{T}^2}$ is conserved, we obtain the homogeneous
solution $T_{\rm cl}(x^0)$
\begin{eqnarray}
  x^0 = \int^{T_{\rm cl}}_0 \frac{dT}{\sqrt{1-V(T)^2/{\cal E}^2}}.
\label{solT}
\end{eqnarray}
When the tachyon approaches its minimum, $V(T) \rightarrow 0$,
the time-dependence of the tachyon simplifies to $T \sim x^0$.
This constant behavior characterizes the final state of the
rolling tachyon.

The location of a static domain wall is determined by the equation
$T_{\rm cl}(x^\mu)=0$ where $T_{\rm cl}$ is the classical
solution of the domain wall, so time-dependent tachyon solutions
are analogously characterized by $T=0$ when the tachyon passes the
top of the potential; the S-brane is found wherever $T=0$. We now
see that the physical statement of (\ref{solT}) is that we have
chosen the S$p$-brane tachyon solution to be the spacelike
$p\!+\!1$-dimensional space $x^0=0$.

Deformations of the S-brane worldvolume are given by analyzing
fluctuations of the tachyon field around its classical solution
(\ref{solT}), $T = T_{\rm cl}(x^0) + t(x^\mu)$. Substituting this
into the action (\ref{senac}) and keeping terms quadratic in $t$,
one is led to the fluctuation action
\begin{eqnarray}
  S_{\rm f} = \frac{-{\cal E}}{2}\!
\int\! dx^0 d^{p+1}x^{\hat{\mu}}
\left(
\frac{-{\cal E}^2}{V^2}\!
\left(\dot{t}\right)^2 \!+\! (\p_{\hat{\mu}} t)^2
\!+\! M^2(x^0)t^2
\right),
\nonumber
\end{eqnarray}
where $\hat{\mu}=1,2,\cdots,p\!+\!1$,
and the time-dependent mass is
\begin{eqnarray}
M^2(x^0)=\left[\frac{V''}{V}- \frac{(V')^2}{V^2}
\right]_{T=T_{\rm cl}}.
\label{mass}
\end{eqnarray}
The factor in front of $\left(\dot{t}\right)^2$ in the fluctuation
action diverges at late time $x^0 \rightarrow \infty$, which
shows that the fluctuation $t$ is governed by the Carrollian
metric \cite{GHY} and ceases to propagate. This is consistent
with the expectation that at the true vacuum of the tachyon
theory open string degrees of freedom disappear and we therefore
concentrate on the fluctuations around S-branes. Since the
solution (\ref{solT}) breaks translation invariance along the
time direction, there is a zero mode on the defect S-brane. It is
well-known that this mode is given by
\begin{eqnarray}
t(x^\mu) = - X^0(x^{\hat{\mu}}) \dot{T}_{\rm cl}(x^0),
\end{eqnarray}
with the function $X^0$ depending only on the coordinates along
the S$p$-brane. Once this is substituted into the fluctuation
action, one finds that the mass term (\ref{mass}) cancels with the
contribution from the term $(\dot{t})^2$. The effective action
for a massless displacement field $X^0(x^{\hat{\mu}})$ is
\begin{eqnarray}
  S= -{\cal T}({\cal E})\int\! d^{p+1}x^{\hat{\mu}}
\; \frac12 (\partial_{\hat{\nu}} X^0)^2,
\end{eqnarray}
with the positive constant ${\cal T}$ depending only on the energy
${\cal E}$.  We have therefore determined the S-brane effective
action for a Euclidean worldvolume to lowest order.

While in the above argument we have introduced only the tachyon
field and its fluctuation $X^0$, it is natural to expect gauge
fields on the S-branes, just like on D-branes. Following the
procedures developed in Ref.\ \cite{tsey}, we shall obtain corrections
to the S-brane action from higher order terms in $X^0$, and
determine couplings to the gauge field in the limit of slowly
varying fields, $\p\p X^0 \sim \p F \sim 0$.  First we note that
the constant gauge field strength appears in the tachyon action
only through the overall Born-Infeld factor $\sqrt{-\det (\eta +
F)_{\mu\nu}}$ and the open string metric $G^{\mu\nu} = ((\eta +
F)^{-1}_{\rm sym})^{\mu\nu}$ used for contracting the
indices of the derivatives (we work in the units
$2\pi\alpha'=1$). Requiring that the equations of motion for the
gauge fields are also satisfied in the time-dependent homogeneous
tachyon background, we find that the open string metric should
satisfy
 $G^{00}\!=\!-1, G^{0\hat{\mu}}\!=\!0$.
This condition, allowing us to introduce dynamical gauge fields
while also preserving the tachyon equations of motion,
essentially states that we can not turn on electric fields on a
Euclidean worldvolume. The second notable point is that the
dependence on the zero mode $X^0$ in the tachyon action should be
\begin{eqnarray}
  S = 
\int\! dx^0d^{p+1}x \; L\left(T_{\rm cl}
\left(\frac{x^0-X^0(x^{\hat{\mu}})}{\beta(X^0)}\right)\right).
\label{betas}
\end{eqnarray}
Here $\beta(X^0)$ can be fixed by the global Lorentz invariance
in the world volume spacetime. The condition that the Lorentz
boost preserves the open string metric is $ \Lambda_\mu^{\;\;\nu}
G_{\nu\rho} (\Lambda^t)^\rho_{\;\;\sigma}= G_{\mu\sigma},$ where
according to the property (\ref{betas}) we define the Lorentz
boost as
\begin{equation}
 \Lambda_\mu^{\ \ \nu}=
\left(
\begin{array}{cc}
1/\beta & -\p_{\hat{\mu}} X^0/\beta \\
 {}* & *
\end{array}
\right).
\end{equation}
Then it is determined that $ \beta =
\sqrt{1-G^{\hat{\mu}\hat{\nu}}\p_{\hat{\mu}} X^0 \p_{\hat {\nu}}
X^0}$. Performing the integration over $x^0$ in Eq.\ (\ref{betas})
and including the $F$-dependence, we obtain the S-brane action
\begin{eqnarray}
 S &=& S_0 ({\cal E})\int\! d^{p+1}x \;\beta(X^0)\sqrt{\det(
\delta + F)_{\hat{\mu}\hat{\nu}}}
\nonumber \\
&=& S_0({\cal E})
\int\! d^{p+1}x \sqrt{\det (\delta_{\hat{\mu}\hat{\nu}}
-\p_{\hat{\mu}} X^0 \p_{\hat{\nu}}X^0 + F_{\hat{\mu}\hat{\nu}})}.
\label{sac}
\end{eqnarray}
The factor $S_0({\cal E})$ is the tachyon action evaluated with
the classical solution $T_{\rm cl}$. The S-brane action
(\ref{sac}) differs from the usual D-brane action (DBI action) in
two important respects: first, the action is defined on a
Euclidean worldvolume, and second the kinetic term of the
transverse scalar field $X^0$ has a ``wrong'' sign since it
represents time translation.
Covariantizing this action, one easily sees that the lagrangian is
simply $\sqrt{\mbox{det}(g+F)}$, where $g$ is the induced metric
on the brane. It differs from the usual DBI lagrangian only by a
factor of $i$, and thus has the same equations of motion. This
ensures that their solutions correspond to consistent backgrounds
in which conformal symmetry is preserved on the open string
worldsheet \cite{Leigh}.


\section{Fundamental strings from S-branes}

The location of the S-brane, given by $X^0(x^{\hat{\mu}})$,
corresponds to the time when the tachyon passes its potential
maximum $T=0$.  So while in the previous section we have shown
that S-branes play an essential role in time-dependent tachyon
condensation, it seems rather surprising if S-branes can be used
to describe remnants of tachyon condensation, especially in the
background $T \rightarrow \infty$. To understand how this is
possible, let us recall how D-branes are realized in the
noncommutative tachyon setup \cite{harvey}. Those remnants
(although in static condensation) are constructed as tachyon
lumps at whose core the tachyon sits still around its potential
maximum ($T\sim 0$) which we can describe with S-branes.
The second obstacle, which is more intrinsically related to this
paper, is that an S-brane is spacelike while any remnants, such
as fundamental strings, are timelike. How do we obtain timelike
objects from spacelike objects? To see this in detail, let us
return to the S-brane action (\ref{sac}). When there are no gauge
fields excited on the S-brane, the lagrangian is $\sqrt{1-(\nabla
X^0)^2}$.  To keep this action real we enforce the condition
$|\nabla X^0| \leq 1$, in units where $c=1$, which is the
statement that excitations on the S-brane keep the worldvolume
Euclidean in the target space. This is analogous to the original
motivation for introducing nonlinear electromagnetism by Born and
Infeld \cite{BI}, if one regards $X^0$ as an electrostatic
potential $\phi$. However, we know that for the DBI action, this
critical electric field can be exceeded by turning on appropriate
fields.  For example turning on transverse scalars $\Phi$ we can
obtain BIon solutions \cite{Gibbons} which have an electric field
exceeding the critical value $1/(2\pi\alpha^\prime)$. Precisely
the same situation can occur for S-branes. Turning on appropriate
magnetic fields on the S-brane allows $|\nabla X^0|>1$, and in
particular, configurations in which $X^0$ goes to infinity. The
physical meaning of having $X^0$ going to infinity is that the
worldvolume of the S-brane exists for all the time. Equivalently
we can say that the S-brane has decayed into branes or strings at
late times.

\begin{figure}[t]
\begin{center}
\begin{minipage}{15cm}
\begin{center}
\epsfxsize=14cm \leavevmode\epsfbox{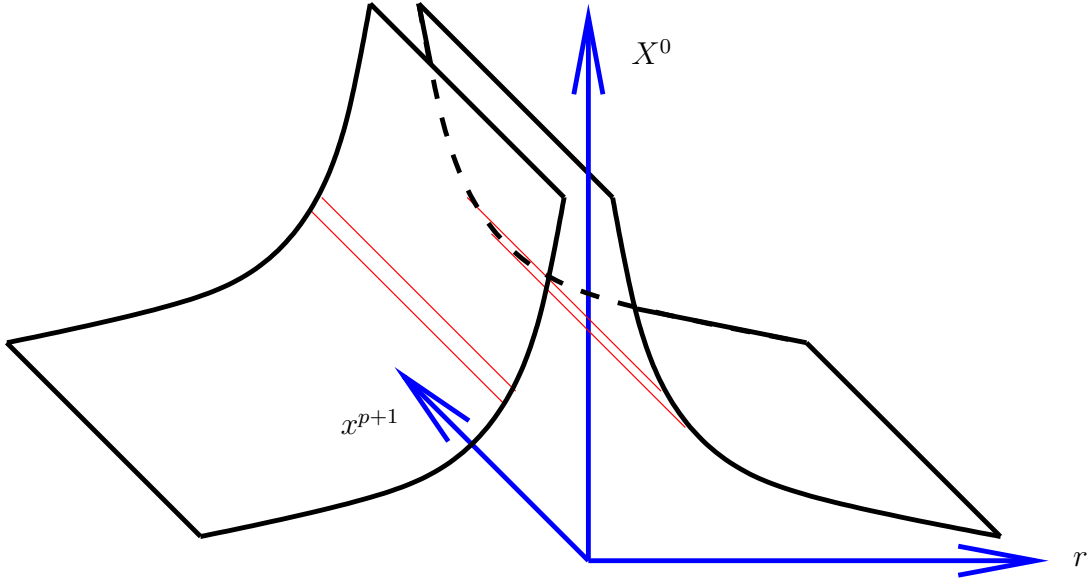} \put(7,0){$r$}
\put(-160,190){$X^0$} \put(-270,50){$x^{p+1}$} \put(0,-10){}
\caption{Formation of a fundamental string as deformation of an
S$p$-brane surface, Eq.~(\ref{spikesol}). The thin double lines
denote the critical time $X^0=X^0_c$ when the spacelike worldvolume
becomes timelike.} \label{spike_fig}
\end{center}
\end{minipage}
\end{center}
\end{figure}

We now discus the BIon-type spike solutions which will represent
how fundamental strings appear as remnants in the tachyon
condensation process. Let us turn on a single gauge potential
$A_{p+1}$ and suppose that all the worldvolume fields are
independent of $x^{p+1}$. Then the S-brane lagrangian is
rewritten as
\begin{eqnarray}
\sqrt{\det (\delta_{\hat{\mu}\hat{\nu}} -\p_{\hat{\mu}} X^0
\p_{\hat{\nu}}X^0 +\p_{\hat{\mu}} A_{p+1} \p_{\hat{\nu}}A_{p+1} )}
\nonumber
\end{eqnarray}
which is exactly the usual static DBI lagrangian for a D$p$-brane,
under the replacement $(X^0,A_{p+1}) \leftrightarrow (\phi,
\Phi)$. Consequently, there are S$p$-brane spike solutions (for
$p\geq 3$) similar to the BPS BIons,
\begin{eqnarray}
  X^0 = A_{p+1} = \frac{c_p}{r^{p-2}} \ .
\label{spikesol}
\end{eqnarray}
Here $r=\sqrt{(x^1)^2 + \cdots + (x^p)^2}$ is the radial distance
along the Euclidean worldvolume (except $x^{p+1}$), but we also see from
(\ref{spikesol}) that $r$ parametrizes time evolution for this
S-brane.  As time evolves the radius decreases, therefore the
remnant becomes a $1+1$ dimensional object, a string,
parametrized by $X^0$ and $x^{p+1}$ (see Fig.\ \ref{spike_fig}).

One can calculate the induced metric
\begin{eqnarray}
ds^2=
(dx^{p+1})^2+
\left(1 - (dX^0/dr)^2\right) dr^2 + r^2 d\Omega_{p-1}^2
\label{induce}
\end{eqnarray}
and find that it is Euclidean for $0<X^0<X^0_c\equiv
c_p^{1/(p\!-\!1)}(p\!-\!2)^{(2\!-\!p)/(p\!-\!1)}$. It is amazing
that for $X_c^0<X^0$ the worldvolume becomes timelike so we are
describing an object which is moving slower than the speed of
light! This S-brane describes an infinitely long cylindrical
worldvolume, where $x^{p+1}$ is the infinite direction and the
radius of the cylinder, $r$, shrinks with time. This therefore
gives the tantalizing possibility that the solution
(\ref{spikesol}) represents a time-dependent tachyon process
which produces fundamental strings.
%
%

Let us provide evidence showing that this remnant is a
fundamental string. First, the field strength induced on the
deformed worldvolume at late times is
\begin{eqnarray}
E_{p+1} =  F_{0 p\!+\!1} = \frac{\partial r}{\partial X^0}F_{r
p\!+\!1} = \frac{1}{2\pi\alpha'}.
\end{eqnarray}
Here we changed the worldvolume variable $r$ to $X^0$, and put in
the $\alpha'$ dependence. This is the critical electric field
induced on the cylindrical worldvolume and indicates that
fundamental string charge is induced on the deformed S-brane. The
criticality of the induced electric field is actually expected in
tachyon condensation on general grounds. In Ref.\ \cite{GHY}, it is
shown that for general tachyon effective lagrangians with
electric field, the late time behavior of the rolling tachyon is
governed by $\dot{T}^2 + E^2 =1$.  When the tachyon stops rolling
the electric field $E$ reaches its critical value.  Since the
S-brane surface is characterized by $T=0$, and at late times the
worldvolume of the S-brane stops shrinking, we find
$\dot{T}\rightarrow 0$. This shows that at late times the domain
$T=0$ is supported only locally at $r=0$ where the electric field
becomes critical. This is consistent with the picture given in
Ref.\ \cite{GHY} for describing fundamental strings after the tachyon
condensation.

Next, we demonstrate that this flux tube has fundamental string
tension. In coordinates more appropriate to the spacetime point
of view, the S-brane action is
\begin{equation}
S= S_0
\int\! dx^0d^{p} x \sqrt{-1 + E_{p+1}^2 +
\dot{r}^2},
\end{equation}
which is $i$ times usual DBI action.
The canonically conjugate momenta are
\begin{eqnarray}
D  =  S_0 \frac{
E}{\sqrt{-1+E^2 + \dot{r}^2}}, \quad
P_r  =  S_0  \frac{\dot{r}}{\sqrt{-1+E^2 +
\dot{r}^2}},
\end{eqnarray}
and the hamiltonian density is
\begin{equation}
H =\frac{S_0}{\sqrt{-1+ E^2 +
\dot{r}^2}} =  \frac{D}{E} \ .
\end{equation}
Imposing flux quantization
$\int\! d^{p-1} x \; D = n$
and noting that the electric field takes the critical
value $E =1$ at late times, we find that the
energy becomes
\begin{equation}
\int\! d^px\;  H = \frac{n}{2\pi\alpha^\prime} \int\! dx^{p+1},
\end{equation}
where we put in the $\alpha'$ dependence. This reproduces the
hamiltonian of $n$ static fundamental strings with fundamental
string tension. The situation resembles that of
Ref.~\cite{nothing} and supertubes \cite{tube} which have critical
electric fields.

Finally, we show that this remnant has no D-brane charge at
future infinity. It is natural to take the Ramond-Ramond (RR)
coupling for an S-brane to be the same as that for a D-brane. The
coupling of RR fields to the particular S-brane above is
\begin{equation}
\mu \int A = \mu \int\! A^{p\!+\!1\, r\,  \Omega}\, r^{p-1} dr\,
dx^{p+1}\, d\Omega_{p-1}.
\end{equation}
Transforming $r$ into the embedding time $X^0$ we obtain
\begin{equation}
\mu \int A^{p\!+\!1\,0\, \Omega}
\left(\frac{X^0}{c_p}\right)^{-\frac{p-1}{p-2}} dX^0\, dx^{p+1}\,
d\Omega_{p-1} \ .
\end{equation}
Due to the factor $(X^0)^{-(p-1)/(p-2)}$ which goes to zero at
late times, we see that the D-brane charge of this solution
shrinks to zero at future infinity.

If as suggested from the static cases \cite{hirano}, we assume
that any S-brane solution has a corresponding tachyon solution,
it is interesting to see why timelike defects are generated
from spacelike defects from the viewpoint of tachyon solutions.
Recall that S-branes are constructed with the open string tachyon
and that the open string light cone always lies inside the closed
string light cone \cite{gib}.
Therefore the region near the spike core $r\sim 0$ can be
timelike (\ref{induce}) in the target space, and also spacelike in
the sense of open string metric due to the diverging field
strength.

T-duality provides another viewpoint of the problem. In the
T-dual picture by compactifying $x^{p+1}$, $A_{p+1}$ becomes a
spatial coordinate, and the S-brane solution (\ref{spikesol}) has
no time-like region. The emergence of the time-like domain in the
original description is simply due to the general fact that
projection of a space-like trajectory onto a lower dimensional
subspace can appear to be time-like.

We have shown that the S-brane can be used to study tachyon
condensation, however it is not clear if the validity of the
S-brane action breaks down at some point due to effects such as
radiation of gravitons, open strings or other modes
\cite{sonogo,BP}.  This point is being investigated.


\section{Conclusions}

The essential point studied in this paper is that one can make
S-branes timelike. Although by definition S-branes are spacelike
objects, they are however constructed using the open string
tachyon and hence governed by the open string metric.  Since the
light cone of this metric always lies inside the closed string
cone, one can boost S-branes to make them timelike relative to the
closed string metric. The construction presented here opens up
new possibilities in describing time-dependent tachyon
condensation with S-branes. Believing that there is
correspondence between the solution of the original tachyon
system and the solution of the defect effective action
\cite{hirano}, the spike solution found in this paper shows how
the fundamental string is formed during time-dependent tachyon
condensation.  The relation of this process to the confinement
picture of the tachyon system \cite{conf} is to be investigated.

In addition, there are numerous avenues to explore using S-brane
actions.  D-brane actions have been truly useful for
investigating string/M theory, and it is intriguing to see what
will be the result of just replacing D-branes by S-branes.
Starting with various brane configurations such as S-brane
junctions and spherical S-branes, one may examine their
supersymmetric properties, their M-theory lift, non-Abelian
S-brane actions, Matrix theory with S-instantons, noncommutative
S-branes, S and T-duality on S-brane actions, spacelike
fundamental strings, S-branes in cubic string field theory, and
numerous subjects. Some of them will be studied in a forthcoming
paper.

Acknowledgements: We thank Chiang-Mei Chen and Miao Li
for useful discussions.
K.\ H.\ is supported in part by
the Grant-in-Aid for Scientific Research No.\ 13135205
from the Japan Ministry of Education, Science and Culture.
The work of P.\ M.\ H.\ and J.\ E.\ W.\
is supported in part by
the National Science Council,
the Center for Theoretical Physics at National Taiwan University,
the National Center for Theoretical Sciences,
and the CosPA project of the Ministry of Education, Taiwan, R.O.C.
The work of P.\ M.\ H. is also supported in part by
the Young Researcher Award of Academia Sinica.

\newcommand{\J}[4]{{\sl #1} {\bf #2} (#3) #4}
\newcommand{\andJ}[3]{{\bf #1} (#2) #3}
\newcommand{\AP}{Ann.\ Phys.\ (N.Y.)}
\newcommand{\MPL}{Mod.\ Phys.\ Lett.}
\newcommand{\NP}{Nucl.\ Phys.}
\newcommand{\PL}{Phys.\ Lett.}
\newcommand{\PR}{ Phys.\ Rev.}
\newcommand{\PRL}{Phys.\ Rev.\ Lett.}
\newcommand{\PTP}{Prog.\ Theor.\ Phys.}
\newcommand{\hep}[1]{{\tt hep-th/{#1}}}


\begin{thebibliography}{10}

\bibitem{stro}
M.\ Gutperle and A.\ Strominger,
{\sl ``Spacelike Branes,''}
\J{JHEP}{0204}{2002}{018}, {\tt hep-th/0202210}.

\bibitem{roll}
A.\ Sen,
{\sl ``Rolling Tachyon,''}
\J{JHEP}{0204}{2002}{048}, {\tt hep-th/0203211};
{\sl ``Tachyon Matter,''}
{\tt hep-th/0203265};
{\sl ``Field Theory of Tachyon Matter,''}
\J{\MPL}{A17}{2002}{1797}, {\tt hep-th/0204143}.

\bibitem{sonogo}
C.\ -M.\ Chen, D.\ V.\ Gal'tsov and M.\ Gutperle,
{\sl ``S-brane Solutions in Supergravity Theories,''}
\J{\PR}{D66}{2002}{024043}, {\tt hep-th/0204071};\\
M.\ Kruczenski, R.\ C.\ Myers and A.\ W.\ Peet,
{\sl ``Supergravity S-Branes,''}
\J{JHEP}{0205}{2002}{039}, {\tt hep-th/0204144};\\
S.\ Roy, 
 {\sl ``On supergravity solutions of space-like Dp-branes,''}
\J{JHEP}{0208}{2002}{025}, {\tt hep-th/0205198};\\
N.\ S.\ Deger and A.\ Kaya,
{\sl ``Intersecting S-Brane Solutions of D=11 Supergravity,''}
\J{JHEP}{0207}{2002}{038}, {\tt hep-th/0206057};\\
K.\ Ohta and T.\ Yokono,
{\sl ``Gravitational Approach to Tachyon Matter,''}
{\tt hep-th/0207004};\\
J.\ E.\ Wang,
{\sl ``Spacelike and Time Dependent Branes from DBI,''}
\J{JHEP}{0210}{2002}{037}, {\tt hep-th/0207089};\\
A.\ Buchel, P.\ Langfelder and J.\ Walcher,
{\sl ``Does the Tachyon Matter?''}
{\tt hep-th/0207235};\\
V.\ D.\ Ivashchuk,
{\sl ``Composite S-brane solutions related to Toda-type systems,''}
{\tt hep-th/0208101};\\
T. Okuda and S. Sugimoto,
{\sl ``Coupling of Rolling Tachyon to Closed Strings,''}
 {\tt hep-th/0208196};\\
A.\ Strominger,
{\sl ``Open String Creation by S-Branes,''}
{\tt hep-th/0209090};\\
B.\ Chen, M. Li and F-L Lin,
{\sl ``Gravitational Radiation of Rolling Tachyon,''}
{\tt hep-th/0209222}.


\bibitem{tsey}
G.\ Arutyunov, S.\ Frolov, S.\ Theisen, A.\ A.\ Tseytlin
{\sl ``Tachyon condensation and universality of DBI action,''}
\J{JHEP}{0102}{2001}{002}, {\tt hep-th/0012080}.

\bibitem{tact}
M.\ R.\ Garousi,
{\sl ``Tachyon couplings on non-BPS D-branes and Dirac-Born-Infeld
 action,''}
\J{\NP}{B584}{2000}{284}, {\tt hep-th/0003122};\\
E.\ A.\ Bergshoeff, M.\ de Roo, T.\ C.\ de Wit, E.\ Eyras and S.\
Panda,
{\sl ``T-duality and Actions for Non-BPS D-branes,''}
\J{JHEP}{0005}{2000}{009}, {\tt hep-th/0003221}.

\bibitem{GHY}
G.\ W.\ Gibbons, K.\ Hashimoto and P.\ Yi,
{\sl ``Tachyon Condensates, Carrollian Contraction of Lorentz Group,
 and Fundamental Strings,''}
\J{JHEP}{0209}{2002}{061}, {\tt hep-th/0209034}.

\bibitem{Leigh}
R.\ G.\ Leigh,
{\sl ``Dirac-Born-Infeld Action From Dirichlet Sigma Model,''}
\J{Mod.\ Phys.\ Lett.}{A4}{1989}{2767}.

\bibitem{harvey}
J.\ A.\ Harvey, P.\ Kraus, F.\ Larsen and E.\ J.\ Martinec,
{\sl ``D-branes and Strings as Non-commutative Solitons,''}
\J{JHEP}{0007}{2000}{042}, {\tt  hep-th/0005031}.

\bibitem{BI}
M.\ Born and L.\ Infeld,
{\sl ``Foundations Of The New Field Theory,''}
\J{Proc.\ Roy.\ Soc.\ London,}{A144}{1934}{425}.

\bibitem{Gibbons}
C.\ G.\ Callan, Jr.\ and J.\ M.\ Maldacena,
{\sl ``Brane Dynamics From the Born-Infeld Action,''}
\J{\NP}{B513}{1998}{198}, {\tt hep-th/9708147};\\
G.\ W.\ Gibbons,
{\sl ``Born-Infeld particles and Dirichlet p-branes,''}
\J{\NP}{B514}{1998}{603}, {\tt hep-th/9709027}.

\bibitem{nothing}
M.\ Kleban, A.\ Lawrence and S.\ Shenker, {\sl ``Closed strings
from nothing,''} \J{\PR}{D64}{2001}{066002}, {\tt hep-th/0012081}.

\bibitem{tube}
D.\ Mateos and P.\ K.\ Townsend,
{\sl ``Supertubes,''}
\J{\PRL}{87}{2001}{011602}, {\tt  hep-th/0103030}.

\bibitem{hirano}
K.\ Hashimoto and S.\ Hirano,
{\sl ``Branes ending on branes in a tachyon model,''}
\J{JHEP}{0104}{2001}{003}, {\tt hep-th/0102173};
{\sl ``Metamorphosis of tachyon profile in unstable D9-branes,''}
\J{\PR}{D65}{2002}{026006}, {\tt hep-th/0102174}.

\bibitem{gib}
G.\ W.\ Gibbons and C.\ Herdeiro,
{\sl ``Born-Infeld Theory and Stringy Causality,''}
\J{\PR}{D63}{2001}{064006}, {\tt hep-th/0008052}.

\bibitem{BP}
C.\ Bachas and M.\ Porrati,
{\sl ``Pair Creation Of Open Strings In An Electric Field,''}
\J{\PL}{B296}{1992}{77}, {\tt hep-th/9209032}.

\bibitem{conf}
O.\ Bergman, K.\ Hori and P.\ Yi, {\sl ``Confinement on the
Brane,''} \J{\NP}{B580}{2000}{289}, {\tt hep-th/0002223};\\
P.\ Yi, {\sl ``Membranes from Fives-Branes and Fundamental Strings
from Dp Branes,''} \J{\NP}{B550}{1999}{214}, {\tt hep-th/9901159};\\
G.\ Gibbons, K.\ Hori and P.\ Yi, {\sl ``String Fluid from
Unstable D-branes,''}
\J{\NP}{B596}{2001}{136}, {\tt hep-th/0009061};\\
A.\ Sen, {\sl ``Fundamental Strings in Open String Theory at the
Tachyonic Vacuum,''}
\J{J.\ Math.\ Phys.}{42}{2001}{2844}, {\tt
hep-th/0010240}.



\end{thebibliography}
\end{document}